\documentclass[pre,showpacs,superscriptaddress,twocolumn]{revtex4-1}
\usepackage{natbib}
\usepackage{amsmath}
\usepackage{amsfonts}
\usepackage{amssymb}
\usepackage{graphicx}

\begin{document}

\title{Shear-induced Casimir forces in liquid layers}

\author{J. M. Ortiz de Z\'arate}
\email{jmortizz@ucm.es}
\affiliation{Departamento de Estructura de la Materia, Facultad de F\'{i}sica, Universidad Complutense, 28040 Madrid, Spain}

\author{T. R. Kirkpatrick}
\email{tedkirkp@umd.edu}
\affiliation{Institute for Physical Science and Technology, University of Maryland, College Park, Maryland 20877, USA}

\author{J. V. Sengers}
\email{sengers@umd.edu}
\affiliation{Institute for Physical Science and Technology, University of Maryland, College Park, Maryland 20877, USA}

\date{\today}

\begin{abstract}
In stationary nonequilibrium states a coupling between hydrodynamic modes causes thermal fluctuations to become long ranged inducing nonequilibrium Casimir forces or pressures. Here we consider nonequilibrium Casimir pressures induced in liquids by a velocity gradient. Specifically, we have obtained explicit expressions for the magnitude of the shear-induced pressure enhancement in a liquid layer between two horizontal plates that complete and correct results previously presented in the literature. In contrast to nonequiibrium Casimir pressures induced by a temperature gradient, kinetic theory shows that nonequilibrium contributions from short-range fluctuations are no longer negligible. In addition, it is noted that computer simulations of model fluids in shear observe effects from molecular correlations at nanoscales that have a different physical origin.  The idea that such computer simulations probe shear-induced pressures resulting from coupling of long-wavelength hydrodynamic modes is erroneous.
\end{abstract}


\maketitle

\section{Introduction}

When large and long-range fluctuations are present, they will induce forces in confined fluids~\cite{KardarGolestanian}. They are commonly referred to as Casimir-like forces in analogy to forces induced by vacuum fluctuations between two conducting plates~\cite{KlimchitskayaEtAl}. Well-known examples are Casimir forces due to critical fluctuations~\cite{KrechBook} or due to long-range correlations in condensed systems with Goldstone modes~\cite{KardarGolestanian}. It has now been well established that even longer-range thermal fluctuations exist in fluids in nonequilibrium states~\cite{DorfmanKirkpatrickSengers}. The physical reason is that the presence of a gradient breaks the symmetry and causes a coupling between long-wavelength hydrodynamic modes~\cite{book}.

In this article we consider Casimir forces due to long-range thermal velocity fluctuations~\cite{LD85,LD02,miJNNFM}. For the case of a liquid layer subjected to a stationary velocity gradient between two parallel plates, we have obtained explicit expressions for the shear-induced pressure enhancement $\delta{p}$, which complete and correct results obtained by previous investigators~\cite{KawasakiGunton,YamadaKawasaki,ErnstEtAl78,WS03}. We provide quantitative estimates for the magnitude of these shear-induced Casimir pressures. In addition, we present an extended kinetic theory approach to compare nonequilibrium Casimir pressures induced by long-range thermal fluctuations with nonequilibrium pressures resulting from short-range thermal fluctuations. We clarify an essential difference between the Casimir pressures caused by macroscopic long-range fluctuations and pressures resulting from fluctuations at nanoscales which are observed in computer simulations~\cite{Evans81,Lee93,Lee94,Marcelli01,Marcelli01B,Ge03,Ahmed10}.

\section{Fluctuation-induced pressures in a liquid under steady shear\label{S2}}

To maintain consistency with a previous analysis of nonequilibrium velocity fluctuations caused by the presence of a velocity gradient~\cite{miCouette,miCouette2}, we continue to use a coordinate system where the $x$ coordinate is in the stream-wise direction, the \textit{y} coordinate in the span-wise direction, and the \textit{z} coordinate in the wall-normal direction. The liquid layer is confined between two horizontal plates located at $z = \pm L$ moving with constant velocities $\pm U$ in the ${x}$ direction. We assume no-slip boundary conditions for the velocities at $z= \pm L$~\cite{Chandra}. The local fluid velocity can be decomposed as $\mathbf{v}={\mathbf{v}}_0(z)+\delta\mathbf{v}$, where ${\boldsymbol{\mathrm{v}}}_0=\{\gamma z,0,0\}$ is the average velocity dependent on the shear rate $\gamma =U/L$ with a  component only in the stream-wise direction \textit{x}, and where $\delta\mathbf{v}(\mathbf{r},t)$ is a fluctuating-velocity contribution dependent on location $\mathbf{r}\left(x,y,z\right)$ and on time \textit{t}. Just as in our previous work on a pressure enhancement induced by a temperature gradient~\cite{miPRL2}, we use nonlinear fluctuating hydrodynamics. Since density fluctuations have been shown to decay faster than velocity fluctuations~\cite{LD02}, the leading fluctuation renormalization to the pressure tensor is (see Appendix~\ref{A3}):
\begin{equation} \label{GrindEQ__1_}
\delta \mathrm{P}\left(\mathbf{r}\right)=\rho {\left\langle \delta \boldsymbol{\mathrm{v}}\mathrm{(}\boldsymbol{\mathrm{r}}\mathrm{)}\delta \boldsymbol{\mathrm{v}}\mathrm{(}\boldsymbol{\mathrm{r}}\mathrm{)}\right\rangle }_{\mathrm{NE}}.
\end{equation}
Here $\rho $ is the mass density and the average is taken over the stationary nonequilibrium (NE) state which is independent of the time \textit{t}. The  diagonal elements $\delta p_{ii\ }=\rho\langle \delta v_i\delta v_i\rangle$ contribute to the shear-induced pressure enhancement, such that $\delta p=\ \frac{1}{3}\left(\delta p_{xx}+\delta p_{yy}+\delta p_{zz}\right).$ The off-diagonal elements all vanish except for $\delta p_{xz}=\ \rho \left\langle \delta v_x\delta v_z\right\rangle $, yielding a fluctuation-induced  contribution to the shear viscosity $\eta$~\cite{ErnstEtAl78}. We find that all diagonal elements depend on the shear rate $\gamma$ and on the Reynolds number $\text{Re}=\gamma L^2/\nu$ as
\begin{equation} \label{GrindEQ__2_}
\delta p_{ii}=V^{\infty }_{ii}~k_BT{\left(\frac{\gamma }{\nu }\right)}^{3/2}{\varphi }_{ii}\left(\mathrm{Re}\right)\ ,
\end{equation}
where $k_B$ is Boltzmann's constant, \textit{T }the temperature, and $\nu $ the kinematic viscosity. Here $\varphi \left(\mathrm{Re}\right)$ is a crossover function, such that for fixed $\gamma$ and sufficiently large $L$ ($\text{Re}\gg1$) ${{\varphi }_{ii}\left(\mathrm{Re}\right)}$ approaches unity, while for fixed $\gamma$ and small $L$ ($\text{Re}\ll1$)  ${{\varphi }_{ii}\left(\mathrm{Re}\right)}$ approaches ${(V}^0_{ii}/V^{\infty }_{ii}){\left(\mathrm{Re}\right)}^{1/2}$. Specifically, the two limiting cases may be written as
\begin{eqnarray} \label{GrindEQ__3_}
\delta p^{\infty }_{ii}&\equiv \displaystyle\lim_{\mathrm{Re}\gg1} \delta{p}_{ii} = V^{\infty }_{ii}~k_BT{\left(\frac{\gamma }{\nu }\right)}^{3/2},\\
\label{GrindEQ__4_}
\delta p^0_{ii}&\equiv \displaystyle\lim_{\mathrm{Re}\ll 1} \delta p_{ii}= V^0_{ii}~k_BTL{\left(\frac{\gamma }{\nu }\right)}^2.
\end{eqnarray}
We emphasize that throughout this paper we consider decay rates sufficiently small so that the flow is laminar and far away from any hydrodynamic instability.

In previous publications we have shown how the correlation functions for the wall-normal velocity fluctuations and for the wall-normal vorticity fluctuations in laminar flow can be derived by solving the fluctuating Orr-Sommerfeld and Squire equations~\cite{miCouette,miCouette2}. For large $L$ ($\text{Re}\gg1$) we have obtained an exact solution of these fluctuating hydrodynamics equations, since in this limit the solution becomes independent of the boundary conditions. For small $L$  ($\mathrm{Re}\ll1$) the solution is strongly affected by the no-slip boundary conditions for the velocity. In this limit it is difficult to get an exact solution~\cite{miORR,miSquire} and we have settled for an approximate solution in a Galerkin approximation~\cite{miCouette,miCouette2}. Explicit expressions for the diagonal elements of the shear-induced pressure tensor can be directly related to the solutions previously obtained for the wall-normal velocity and vorticity fluctuations as shown in Appendix~\ref{A1} for $\text{Re}\gg1$ and in Appendix~\ref{A2} for $\text{Re}\ll1$. These results are directly obtained from the nonequilibrium correlation functions presented in~\cite{miCouette,miCouette2}; the only additional step required is integration of the correlation functions over wave numbers so as to get the intensity of the velocity fluctuations in real space. For the coefficients in Eqs.~\eqref{GrindEQ__3_} and~\eqref{GrindEQ__4_} we have thus obtained: $V^{\infty }_{xx}=+0.0847,~V^{\infty }_{yy}=+0.0173,\ V^{\infty }_{zz}=+0.0106,\ $and $V^0_{xx}=+0.001438,\ V^0_{yy}=+0.000480,\ V^0_{zz}=+0.000392$.

In the case of $\text{Re}\ll1$  the actual solution of the fluctuating hydrodynamics equations yield expressions for $V^0_{ii}$  that depend on the \textit{z }coordinate as a consequence of the boundary conditions at $z/L=\pm 1$ (see Eq.~\eqref{E4} in Appendix~\ref{A2}). However, just as in the case of the Casimir pressures induced by a temperature gradient~\cite{miPRL2}, mechanical equilibrium, combined with conservation of mass, causes a uniform pressure enhancement equal to the height-averaged value obtained from the fluctuating hydrodynamics equations. Hence, we only quote here the height-averaged values for $V^0_{ii}$.

Upon substituting  the results quoted above for $V_{ii}^\infty$ and $V_{ii}^0$ into Eqs.~\eqref{GrindEQ__3_} and~\eqref{GrindEQ__4_} we conclude:
\begin{eqnarray} \label{GrindEQ__5_}
\delta{p}^{\infty}&=\frac{1}{3}\sum_i{\delta p^{\infty }_{ii}=+0.0375~k_BT{\left(\frac{\gamma }{\nu }\right)}^{3/2}},\\
\label{GrindEQ__6_}
\delta{p}^0&=\frac{1}{3}\sum_i{\delta p^0_{ii}=+0.000770~k_BT{L\left(\frac{\gamma }{\nu }\right)}^2}.
\end{eqnarray}
We note that the anomalous dependence of $\delta p\ $on ${\gamma }^{3/2}$ and on $L{\gamma }^2$ is a consequence of a coupling between macroscopic viscous modes and not from static or dynamic correlations at molecular scales.
We do not consider contributions from the sound modes here, since they are smaller by a factor of ${\left(U/c\right)}^{1/2}$ , where \textit{c} is the speed of sound. Terms of higher order in the shear-rate have also been neglected as discussed in Appendix~\ref{A3}. We note from Eq.~\eqref{GrindEQ__6_} that the shear-induced pressure $\delta{p}^0$ \emph{increases} with $L$ at a constant shear rate $\gamma=U/L$, but \emph{decreases} with $L$ at a constant velocity $U$. This is similar to the fluctuation-induced pressure in a liquid subjected to a temperature gradient that increases with $L$ at a constant  temperature gradient $\nabla{T}= \Delta{T}/L$, but decreases with $L$ at a constant temperature difference $\Delta{T}$ between the plates~\cite{miPRL2}.

\section{Interpretation of long-ranged pressure contributions\label{S3}}

Attempts to determine the shear-induced pressure tensor in the absence of boundary conditions have been made by Kawasaki and Gunton~\cite{KawasakiGunton} and by Yamada and Kawasaki~\cite{YamadaKawasaki}. While they did find that the shear-induced pressure varies with the shear rate as ${\gamma }^{3/2}$, the numerical coefficients are substantially different from the values found by us as shown in Table~\ref{T01}.
\begin{table}
\caption{Comparison with literature}\label{T01}
\begin{tabular*}{\columnwidth}{l@{\extracolsep{\fill}}lccc}
\toprule
&$V^{\infty }_{xx}$&$V^{\infty }_{yy}$&$V^{\infty }_{zz}$  \\[6pt]
\colrule
Kawasaki and Gunton~\cite{KawasakiGunton}   & +0.0050&   --0.0046&     --0.0017\\
Yamada and Kawasaki~\cite{YamadaKawasaki}  &+0.0428 &  +0.0173  &  +0.0106\\
This work                 &+0.0847 &  +0.0046  &  +0.0106\\
\botrule
\end{tabular*}
\end{table}

Ernst \textit{et al.}~\cite{ErnstEtAl78} determined the traceless part of the shear-induced pressure tensor using a kinetic-theory approach. Our results for the traceless part of the shear-induced pressure tensor are in perfect agreement with those obtained by Ernst \textit{et al.} as shown in Table II. In this table we also see perfect agreement with their off-diagonal stress, $V_{xz}$, which gives a generalized viscosity. Hence, we are confident that we have obtained correct expressions for the shear-induced pressure enhancement as given by Eqs. \eqref{GrindEQ__5_} and \eqref{GrindEQ__6_}. Wada and Sasa~\cite{WS03} have only determined the wall-normal component of the shear-induced pressure tensor. They find $\ V^{\infty }_{zz}$ = +0.0106 in agreement with our result, but their value $V^0_{zz}$ = 0.000553 slightly differs from $V^0_{zz}$ = 0.000392 found by us. The reason is that Wada and Sasa used periodic boundary conditions which are mathematically convenient, but physically unrealistic.

\begin{table*}
\caption{Traceless part of shear-induced pressure tensor}\label{T02}
\begin{tabular*}{\textwidth}{l@{\extracolsep{\fill}}lcccc}
\toprule
&$V^{\infty }_{xx}-\frac{1}{3}\sum_i{V^{\infty }_{ii}}$&${V}^{\infty }_{yy}-\frac{1}{3}\sum_i{V^{\infty }_{ii}}$&$V^{\infty }_{zz}-\frac{1}{3}\sum_i{V^{\infty }_{ii}}$&$V^{\infty }_{xz}$\\[6pt]
\colrule
Ernst \textit{et al}.~\cite{ErnstEtAl78}  &+0.0470&--0.0202&--0.0268&+0.00916\\
This work                 &+0.0472&--0.0202&--0.0269&+0.00916\\
\botrule
\end{tabular*}
\end{table*}

To estimate the magnitude of the shear-induced pressure enhancement we consider water, which is the liquid commonly used in Couette-flow experiments~\cite{REF28,REF29,REF30,REF31,REF32,REF33,REF34,REF35}. The smallest gap width thus far employed is about 1.5~mm~\cite{REF32}. The possible experimental plate velocities \textit{U} may be up to $0.5~\text{m s}^{-1 }$~\cite{endnote36}. A gap width of 1 mm (\textit{L }= 0.5 mm) and plate velocities ${U }= \pm  0.5~\text{ms}^{-1}$ imply Re $\approx \ $280, which is still well below the critical Reynolds number for the onset of turbulence~\cite{REF28}. Substituting $\nu =8.93\times {10}^{-7\ }{\mathrm{m}}^2{\mathrm{s}}^{-1}$ for the kinematic viscosity of water at 298.15~K~\cite{endnote37} into Eqs. \eqref{GrindEQ__5_} and \eqref{GrindEQ__6_} we obtain the estimates
\begin{align}\label{GrindEQ__7_}
\delta{p}^{\infty }&=6\times {10}^{-9}~\text{Pa}& \text{and}& &\delta{p}^0&=2\times {10}^{-9}~\text{Pa},
\end{align}
\textit{i.e.,} the shear-induced pressure enhancement is somewhere between ${10}^{-9}$ and ${10}^{-8}$ Pa. It is interesting to compare this shear-induced pressure enhancement with those in a liquid layer with the same gap width either from critical fluctuations $\delta p=-2\times {10}^{-11}\ \mathrm{Pa}$ (from Ref.~\cite{miPRE2015}, corrected for a sign error) or from nonequilibrum temperature fluctuations caused by the presence of a temperature gradient (25 K /mm) $\delta p=5\times {10}^{-4}\ \mathrm{Pa}$~\cite{miPRL2}.

We thus see that the shear-induced pressure enhancement is many orders of magnitude smaller than the Casimir pressures induced by the presence of a temperature gradient. One reason is that temperature fluctuations decay more slowly than velocity fluctuations and, hence, are more strongly impacted by the presence of a temperature gradient. Another reason is that the shear-induced pressure enhancement has a kinetic origin, while the pressure enhancement from a temperature gradient has a potential origin that in liquids is several orders of magnitude larger.

We also note that in the derivation of Eq.~\eqref{GrindEQ__7_} for the shear-induced pressure we have considered isothermal flow. That is, possible viscous heating effects have been neglected~\cite{DL85}. This condition is commonly satisfied in computer simulations by special dynamical rules keeping the temperature constant. However, as we shall show in a subsequent publication, it turns out that in real experiments a pressure increase resulting from viscous heating is not negligible.

\section{Other shear-induced pressure contributions}
\subsection{Short-ranged kinetic contributions}

An important difference, between the giant Casimir pressures  in liquids subjected to a temperature gradient~\cite{miPRL2} and the Casimir pressures  in the presence of shear, is that in the case of shear short-ranged contributions are no longer negligible. To elucidate a possible contribution from short-range correlations, we note from nonequilibrium statistical mechanics that $\delta p=\kappa {\gamma }^2$, where $\kappa $ is a nonlinear Burnett coefficient. These nonlinear Burnett coefficients are known to diverge as $L\to \infty $~\cite{Brey}. We may decompose this Burnett coefficient as the sum of a finite short-range contribution ${\kappa }^{(0)}$ and a long-range contribution $L{\kappa }^{\eqref{GrindEQ__1_}}$~\cite{miPRL2}, yielding a short-range (SR) and a long-range (LR) contribution to the shear-induced pressure enhancement:
\begin{equation} \label{GrindEQ__8_}
\delta p=\delta p_{\mathrm{SR}}+\delta p_{\mathrm{LR}},
\end{equation}
where $\delta p_{\mathrm{SR}}={\kappa }^{(0)}{\gamma }^2$ and $\delta p_{\mathrm{LR}}=L{\kappa }^{\eqref{GrindEQ__1_}}{\gamma }^2$. Comparing with Eq. \eqref{GrindEQ__6_}, we note that the shear-induced Casimir pressure arises from the same long-wavelength hydrodynamic modes that cause the nonlinear Burnett coefficient $\kappa $ to diverge. A complete kinetic theory for the nonlinear Burnett coefficients of real fluids is not available, but it is possible to get an order-of-magnitude estimate for the SR contribution by extending the theory of Enskog for the transport properties of a dense gas of hard spheres to the quadratic level~\cite{VanBeijerenDorfmann}. Starting from an expression for the pressure tensor of a gas of hard spheres provided by Dufty~\cite{Dufty02} and retaining only the collisional transfer contribution, which is the dominant one at high densities, we obtain
\begin{equation} \label{GrindEQ__9_}
\delta p_{\mathrm{SR}}\cong \rho {\sigma }^2n{\sigma }^3\frac{7\pi }{45}\chi {\gamma }^2,
\end{equation}
where $\sigma $ is the hard-sphere diameter,  \textit{n} the number density, and $\chi $ the value of the radial distribution function at contact between the spheres. In principle, there are short-range corrections to Eq.~\eqref{GrindEQ__9_} of $O(\gamma^4)$. In dimensionless variables, these terms are of relative $O([\gamma\sigma/v_{th}]^2)$, with $v_{\mathrm{th}}\propto\sqrt{k_BT/m}$ a thermal velocity. For realistic laboratory shear rates these are very small corrections and can be neglected. Then, since for liquid water $\rho =nm={10}^3\mathrm{\ kg\ }{\mathrm{m}}^{-3}$, $m=3\times {10}^{-26}\ \mathrm{kg}$, $\sigma =3\times {10}^{-10}\ \mathrm{m}$, and estimating $\chi \cong 5$ for a dense liquid,  we then conclude from Eq. \eqref{GrindEQ__9_} that for water with ${L }= 0.5~\text{mm}$ and ${U }= 0.5~ \text{ms}^{-1}$
\begin{equation} \label{GrindEQ__10_}
\delta p_{\mathrm{SR}}\cong \ 2\times {10}^{-10}\mathrm{\ Pa}.
\end{equation}
On comparing Eq. \eqref{GrindEQ__10_} with Eq. \eqref{GrindEQ__7_} we see that the SR contribution to the induced-pressure enhancement is indeed smaller than the LR contribution to the shear-induced pressure enhancement, but may not be entirely negligible, even at \textit{L} = 0.5 mm. The SR contribution becomes even more important at smaller values of \textit{L}. From Eq. \eqref{GrindEQ__9_} it follows that, for a fixed velocity \textit{U},\textit{ }$\delta p_{\mathrm{SR}}$ will increase as \textit{L}$^{-2}$, while $\delta p_{\mathrm{LR}}$, due to the long-range velocity fluctuations, will only increase either as \textit{L}${}^{-3/2 }$ for large values of Re or even less as ${L}^{-1}$ for small values of Re.

\subsection{Computer simulations and nanoscale contributions}

Another related important consequence is that most computer simulations of model fluids under shear have generally been misinterpreted~\cite{Lee93,Lee94,Marcelli01,Marcelli01B,Ge03,Ahmed10}. Investigators have either claimed to have found agreement~\cite{Lee93,Lee94} or disagreement~\cite{Marcelli01,Marcelli01B,Ge03,Ahmed10} with the predictions of Eqs. \eqref{GrindEQ__3_} and \eqref{GrindEQ__4_}. However, these computer simulations probe effects of fluctuations at nanoscales, which have a completely different physical origin and need to be distinguished from the long-range macroscopic fluctuations responsible for the shear-induced Casimir pressure described by Eqs. \eqref{GrindEQ__2_} -- \eqref{GrindEQ__4_}.

The first molecular dynamics (MD) simulations on a 3-dimensional sheared fluid consisting of a small number of Lennard-Jones (LJ) particles were performed by Evans~\cite{Evans81}. He found results that seemed, especially near the triple point, to indicated a nonequilibrium (NE) pressure enhancement that was proportional to ${\gamma }^{3/2}$, but with a coefficient that was orders of magnitude larger than the coefficient to be expected from Eq.~\eqref{GrindEQ__3_}. He noted a similarity with the so-called molasses tail observed in MD simulations of the equilibrium stress-tensor time correlation functions that determines the shear viscosity~\cite{REF42}. In turns out that in this time-dependent correlation function, again near the triple point of LJ particles or near freezing of hard-sphere particles, an apparent long-time tail proportional to $1/t^{3/2}$ appears, but with a coefficient, again, several orders of magnitude larger than the theoretically expected long-time tail coefficient. It was subsequently realized that this molasses tail was not due to long-wave length mode-coupling (MC) effects, but was due to molecular scale MC effects related to structural relaxation in dense fluids~\cite{REF43,REF44,REF45,REF46}. This theory explains the magnitude of observed molasses tails and predicts that this $1/t^{3/2}\ $ behavior will crossover to an exponential decay on a structural time scale ${\tau }_{\mathrm{s}}\approx S(k_0)/Dk^2_0$, where $D$ is the self-diffusion coefficient and $k_0$ the wave number at which the static structure factor $S(k)$ has its maximum~\cite{REF44}. For a review of these molecular scale MC effects, the reader is referred to a forthcoming book of Dorfman \textit{et al.}~\cite{REF47}. Another complication is that the computer simulations use extremely large shear rates $\gamma \approx {10}^{11}-{10}^{12}~{\mathrm{s}}^{-1}$. At such large shear rates, where $\gamma >{\tau }^{-1}_{\mathrm{s}}$, the NE pressure is also determined by molecular-scale MC effects. The molecular-scale effects will not only depend on the intermolecular potential adopted, but, at a given density, also on the number of free paths sampled, and, hence, on the number of particles used in the simulations.

Generally, it never makes sense, except in some asymptotic limit, to fit the shear-induced pressure enhancement in terms of a simple power law. For example, Eq. \eqref{GrindEQ__8_} suggests for sufficiently large \textit{L} a fit to
\begin{equation} \label{GrindEQ__11_}
\delta p=A_{\mathrm{SR}}{\gamma }^2+\delta p_{\mathrm{LR}}
\end{equation}
with $\delta p_{\mathrm{LR}}\propto L{\gamma }^2$ for Re $<$ 1 or $\delta p_{\mathrm{LR}}\propto {\gamma }^{3/2}$ for Re $>$ 1, and with $A_{\mathrm{SR}}$ independent of \textit{L}. In MD simulations that probe structural relaxation effects the appropriate fit should have an additional term $\delta p_{\mathrm{s}}\propto {\gamma }^{3/2}$ for $\gamma >{\tau }^{-1}_{\mathrm{s}}$ and $\delta p_{\mathrm{s}}\propto {\gamma }^2$ for $\gamma <{\tau }^{-1}_{\mathrm{s}}$. That is, for $\gamma >{\tau }^{-1}_{\mathrm{s}}$, $\delta p_{\mathrm{s}}$ will renormalize $\delta p_{\mathrm{LR}}$ in Eq. \eqref{GrindEQ__11_} and for $\gamma <{\tau }^{-1}_{\mathrm{s}}$, $\delta p_{\mathrm{s}}$ will renormalize $A_{\mathrm{SR}}$ in this equation.

Almost all discussions of computer-simulation studies currently available~\cite{Lee93,Lee94,Marcelli01,Marcelli01B,Ge03,Ahmed10} have ignored the effects of the molecular-scale correlations that are dominant at nanoscales. Lee and Cumming~\cite{Lee93,Lee94} found an enhancement $\propto {\gamma }^{3/2}$. But without checking the coefficient, they assumed to have found agreement with both the results of Evans~\cite{Evans81} and with Eq.~\eqref{GrindEQ__3_}, which is impossible as explained above. Sadus and coworkers~\cite{Marcelli01,Marcelli01B,Ge03,Ahmed10} have found effective exponents for the shear-rate dependence ranging from 1.5 to 2 without any theoretical analysis of the results.

The theoretical expression, Eq.~\eqref{GrindEQ__2_}, for the shear-induced pressure enhancement follows from a solution of the fluctuating hydrodynamics equations for the long-range velocity fluctuations. Numerical solutions of the fluctuating hydrodynamics equations have been obtained, some years ago with the direct simulation Monte Carlo method~\cite{Alejandro1,Alejandro2} and, more recently, by Varghese \emph{et al.}~\cite{Varghese17} with a multiparticle collision dynamics method. These approaches apply either to dilute gases~\cite{Alejandro1,Alejandro2} or to a model fluid with an ideal-gas equation of state~\cite{Varghese17}, but have the merit of evaluating the Casimir pressure purely mechanically, from momentum exchange in particle-wall collisions~\cite{Alejandro2}. The most recent work~\cite{Varghese17} concludes that the shear-induced pressure enhancement obtained over about one decade of the shear-rate appears to scale as $\gamma^2$ and therefore does not agree with Eq.~\eqref{GrindEQ__5_}. However, the magnitude of the enhancement seems to be indeed of the order given by our Eqs.~\eqref{GrindEQ__5_} and~\eqref{GrindEQ__6_}. It would be of interest to pursue such recent numerical calculations for a larger ranges of $L$ and Re numbers, so as to probe a possible crossover from a behavior $\propto L\gamma^2$ for small Re to $\propto\gamma^{3/2}$ for large Re. Varghese \emph{et al.}~\cite{Varghese17} conclude their paper with the following comment: ``It therefore remains for further theoretical and simulation studies to establish a unified picture of the exponent associated with the hydrodynamic pressure under shear". This article attempts to address this issue.

\section{Conclusions}

Velocity fluctuations in sheared liquids, resulting from coupling of long-range hydrodynamic modes, induce Casimir pressures in confined liquid layers. However, in contrast to the case of Casimir pressures induced by nonequilibrium temperature fluctuations~\cite{miPRL2}, pressure contributions from nonequilibrium short-range correlations are no longer negligible. Moreover, almost all computer simulations of liquids under shear probe molecular correlations at nanoscales that have a different physical origin.

\section*{Acknowledgements}

We thank J.R. Dorfman for valuable discussions and R. Monchaux for some comments regarding Couette-flow experiments. The research at the Complutense University was supported by grant ESP2017-83544-C3-2-P of the Spanish \emph{Agencia Estatal de Investigaci\'on}. The research at the University of Maryland was supported by the US National Science Foundation under Grant No. DMR-1401449.

\appendix

\section{Other contributions to the shear-induced pressure enhancement\label{A3}}

The long wavelength nonlinear terms in the nonlinear fluctuating hydrodynamic equations that renormalize the pressure are
\begin{equation}
\langle\delta P_{ij}\rangle=\langle\rho v_iv_j+p(\rho,\epsilon)\delta_{ij}-\overline\rho\overline{v}_i\overline{v}_j-p(\overline\rho,\overline\epsilon)\delta_{ij}\rangle.
\end{equation}
Here the over-line denotes average values and $\epsilon$ is the internal-energy density. In principle all these nonlinearities will lead to long-ranged renormalizations of the pressure.
If we neglect density and internal-energy density fluctuations, then we obtain Eq.~(1). Taking into account density fluctuations leads to additional renormalizations of the pressure. However, Lutsko and Dufty~\cite{LD02} have shown that in general density fluctuations decay faster in space than velocity fluctuations. Their work suggest that density nonlinearities will lead to a $\gamma^{11/6}$ contribution to the pressure, with a relatively large coefficient. Compared to the $\gamma^{3/2}$ terms, this term is of relative $O[\gamma\sigma/v_{\mathrm{th}}]^{1/3}$, where $\sigma$ is a molecular size and where $v_\text{th}$ is the thermal velocity; again this term is quite small. In any case, it would be difficult to distinguish this term from all of the analytic $\gamma^2$ terms.

\section{Calculation for $\text{Re}\gg1$\label{A1}}

For the calculations it is convenient to use dimensionless variables with position \textbf{r} in terms of \textit{L, }wave vector \textbf{q }in terms of $L^{-1}$, and velocity $\mathbf{v}$ in  terms of $L\gamma $. Then, all the quantities of interest depend only on the Reynolds number and a dimensionless strength of the thermal noise as~\cite{miCouette}:
\begin{equation}
\tilde{S}=\frac{k_BT}{\rho L^2}\frac{1}{{\gamma }^2L^2}\frac{1}{\mathrm{Re}}.
\end{equation}
Large \textit{L} and small \textit{L} at a fixed shear rate $\gamma$ correspond to $\text{Re}\gg1$ and $\text{Re}\ll1$, respectively. For large \textit{L} we can neglect the boundary conditions and solve the fluctuating Orr-Sommerfeld and Squire equations by applying a 3-dimensional Fourier transformation~\cite{miCouette,miCouette2}. We then obtain for the NE part of the equal-time correlation functions in momentum space:
\begin{eqnarray}\label{ES2}
{\left\langle \delta v^*_z\left(\boldsymbol{\mathrm{q}}\right)\delta v_z\left(\boldsymbol{\mathrm{q}}'\right)\right\rangle }_{\mathrm{NE}}&=&C^{\mathrm{NE}}_{zz}(\mathbf{{q}})~{\left(2\pi \right)}^3~ \delta(\boldsymbol{\mathrm{q}}-\boldsymbol{\mathrm{q}}'),\\
{\left\langle \delta {\omega }^*_z\left(\boldsymbol{\mathrm{q}}\right)\delta {\omega }_z\left(\boldsymbol{\mathrm{q}}'\right)\right\rangle }_{\mathrm{NE}}&=&W^{\mathrm{NE}}_{zz}(\mathbf{{q}})~{\left(2\pi \right)}^3~ \delta(\boldsymbol{\mathrm{q}}-\boldsymbol{\mathrm{q}}'),\\
{\left\langle \delta v^*_z\left(\boldsymbol{\mathrm{q}}\right)\delta {\omega }_z\left(\boldsymbol{\mathrm{q}}'\right)\right\rangle }_{\mathrm{NE}}&=&\mathrm{i}~B^{\mathrm{NE}}_{zz}(\mathbf{q})~{\left(2\pi \right)}^3~ \delta(\boldsymbol{\mathrm{q}}-\boldsymbol{\mathrm{q}}').\label{ES4}
\end{eqnarray}
The functions $C^{\mathrm{NE}}_{zz}\left(\boldsymbol{\mathrm{q}}\right)$, $W^{\mathrm{NE}}_{zz}\left(\boldsymbol{\mathrm{q}}\right)$, and $B^{\mathrm{NE}}_{zz}\left(\boldsymbol{\mathrm{q}}\right)$ are given by
\begin{align}\label{ES5}
\frac{C^{\mathrm{NE}}_{zz}(\mathbf{{q}})}{\tilde{S}\text{Re}}&=2\frac{q_xq^2_{\parallel }}{q^4}\int^{\infty }_0{d\beta \left(q_z+q_x\beta \right)}~e^{-\mathrm{\Gamma }\left(\beta ,\boldsymbol{\mathrm{q}}\right)},\\
\label{ES6}\frac{W^{\mathrm{NE}}_{zz}(\mathbf{{q}})}{\tilde{S}\mathrm{Re}}&=\frac{q_y^2}{q_x^2}\int^{\infty}_0 \hspace*{-10pt}d\beta~\left[\frac{d\Gamma}{d\beta}\right]^2~
\left[U(\beta,\mathbf{q})\right]^2~e^{-\mathrm{\Gamma }\left(\beta ,\boldsymbol{\mathrm{q}}\right)},\\
\label{ES7}
\frac{B_{zz}^\text{NE}(\mathbf{q})}{\tilde{S}\mathrm{Re}}&=\frac{q_\parallel q_y}{q^2 q_x}\int_0^\infty\hspace*{-10pt}d\beta~\left[\frac{d\Gamma}{d\beta}\right]^2~U(\beta,\mathbf{q})~e^{-\Gamma(\beta,\mathbf{q})}.
\end{align}
with
\begin{equation}
\begin{split}
\mathrm{\Gamma }(\beta,\mathbf{q})&=\frac{2\beta }{3\mathrm{Re}}\left(q^2_x{\beta }^2+3\beta q_xq_z+3q^2\right),\\
U(\beta,\mathbf{q})&=\frac{\text{Re}}{2} \left[\text{atan}\left(\frac{q_z+\beta{q_x}}{q_\parallel}\right) - \text{atan}\left(\frac{q_z}{q_\parallel}\right)\right],\\
&=q_xq_\parallel\int_0^\beta \left[{\frac{d\Gamma(u)}{du}}\right]^{-1}~du,
\end{split}
\end{equation}
as given by Eq.~(39) and Eq.~(43b) in ref.~\cite{miCouette2}, which are exactly the same as Eqs.~\eqref{ES5} and~\eqref{ES6} here, while Eq.~\eqref{ES7} for the cross-correlation can be obtained following the same techniques and is first presented here. In these equations ${q}_{\parallel }$ is the magnitude of the component ${\mathbf{q}}_{\parallel }$ of the wave vector in the \textit{x-y} plane, \textit{i.e.,} parallel to the plates. From these equations we obtain the correlation functions for $\delta v_x$ and $\delta v_y$ by noting that
\begin{eqnarray}
\delta v_x=\frac{-1}{q^2_{\parallel }}\left(q_xq_z\delta v_z-\mathrm{i}q_y\delta {\omega }_z\right),\\
\delta v_y=\frac{-1}{q^2_{\parallel }}\left(q_yq_z\delta v_z+\mathrm{i}q_x\delta {\omega }_z\right),
\end{eqnarray}
so that, in conjunction with the divergence-free flow condition, $\boldsymbol\nabla\cdot\delta\mathbf{v}=0$, one has
\begin{equation}\label{ES10}
\begin{split}
{\left\langle \delta v^*_x\left(\boldsymbol{\mathrm{q}}\right)\delta v_x\left(\boldsymbol{\mathrm{q}}'\right)\right\rangle }_{\mathrm{NE}}&=C^{\mathrm{NE}}_{xx}\left(\boldsymbol{\mathrm{q}}\right){\left(2\pi \right)}^3\delta \left(\boldsymbol{\mathrm{q}}-\boldsymbol{\mathrm{q}}'\right),\\
{\left\langle \delta v^*_y\left(\boldsymbol{\mathrm{q}}\right)\delta v_y\left(\boldsymbol{\mathrm{q}}'\right)\right\rangle }_{\mathrm{NE}}&=C^{\mathrm{NE}}_{yy}\left(\boldsymbol{\mathrm{q}}\right){\left(2\pi \right)}^3\delta \left(\boldsymbol{\mathrm{q}}-\boldsymbol{\mathrm{q}}'\right),
\end{split}
\end{equation}
with
\begin{eqnarray}
C^{\mathrm{NE}}_{xx}\left(\boldsymbol{\mathrm{q}}\right)=\frac{q^2_xq^2_z}{q^4_{\parallel }}C^{\mathrm{NE}}_{zz}\left(\mathrm{q}\right)+\frac{q^2_y}{q^4_{\parallel }}W^{\mathrm{NE}}_{zz}\left(\mathrm{q}\right)\notag\\+2\frac{q_xq_yq_z}{q^4_{\parallel }}B^{\mathrm{NE}}_{zz}\left(\mathrm{q}\right),\label{ES13}\\
C^{\mathrm{NE}}_{yy}\left(\boldsymbol{\mathrm{q}}\right)=\frac{q^2_yq^2_z}{q^4_{\parallel }}C^{\mathrm{NE}}_{zz}\left(\mathrm{q}\right)+\frac{q^2_x}{q^4_{\parallel }}W^{\mathrm{NE}}_{zz}\left(\mathrm{q}\right)\notag\\-2\frac{q_xq_yq_z}{q^4_{\parallel }}B^{\mathrm{NE}}_{zz}\left(\mathrm{q}\right).\label{ES14}
\end{eqnarray}
Integration of Eqs.~\eqref{ES5}, \eqref{ES13}, and~\eqref{ES14} yields the diagonal elements of ${\left\langle \delta \boldsymbol{\mathrm{v}}\delta \boldsymbol{\mathrm{v}}\right\rangle }_{\mathrm{NE}}$ in real space for large Re.

As an example, we consider the computation of $V_{zz}^\infty$. In terms of dimensionless units we have:
\begin{equation}\label{ES15}
\tilde{S}\text{Re}~V_{zz}^\infty~(\text{Re})^{3/2} = \frac{1}{(2\pi)^3}\int_{\mathbb{R}^3} C_{zz}^\text{NE}(\mathbf{q})~d\mathbf{q}.
\end{equation}
To evaluate the coefficient $V_{zz}^\infty$, after substitution of Eq.~\eqref{ES5} into Eq.~\eqref{ES15}, we adopt spherical coordinates for the integration over $\mathbf{q}$. We first integrate over the magnitude $q$ of the vector $\mathbf{q}$, which can be done analytically and yields the prefactor $(\text{Re})^{3/2}$. A second integration over the polar angle can also be performed analytically taking advantage of the symmetry properties of the integral. The final double integral, over the azimuthal angle and over the parameter $\beta$, can be simplified but not performed analytically and has been evaluated numerically:
\begin{equation}
\begin{split}
V_{zz}^\infty&=\frac{\sqrt{3}}{32\pi^3}\Gamma(\tfrac{1}{4})^2\int_{0}^\infty\frac{d\beta}{\beta^\frac{3}{2}}\int_0^\pi \frac{(\beta+\cos\theta)~(\sin\theta)^\frac{9}{2}}{(\beta^2+3\beta\cos\theta+3)^\frac{3}{2}}~d\theta\\
&\simeq 0.0106,
\end{split}
\end{equation}
which is the value quoted in Table~\ref{T01} of the main text. The other coefficients, $V_{xx}^\infty$ and $V_{yy}^\infty$, are evaluated in a similar fashion from Eqs.~\eqref{ES13}-\eqref{ES14}. The resulting values are $V^{\infty }_{xx}=+0.0847,~V^{\infty }_{yy}=+0.0173$, as also shown in Table~\ref{T01} where a detailed discussion and comparison with literature is presented.

\section{Calculation for $\text{Re}\ll1$\label{A2}}

Small Re means narrow layers. Hence, we need to take the boundary conditions for the velocity fluctuations at dimensionless $z=\pm 1$ into account explicitly. As shown in~\cite{miCouette,miCouette2}, this is accomplished by applying a Fourier transformation only in the stream-wise and span-wise directions, while retaining the dependence of the wall-normal coordinate $z$. The presence of boundaries breaks the translational invariance in the wall-normal direction and the equal-time correlation functions are invariant only in the $xy$-plane. Therefore, their equal-time Fourier transforms shall be proportional to delta functions $\delta({{\mathbf{q}}}_{{\parallel}}-{{\mathbf{q}}}^{\prime}_{{\parallel}})$ so that they can be expressed most compactly as:
\begin{eqnarray}\label{ES17}
\frac{{\left\langle \delta v^*_z({\mathbf{q}}_{\parallel },z)~\delta v_z({\mathbf{q}}^{\prime}_{\parallel},z^\prime)\right\rangle }_{\mathrm{NE}}}{(2\pi)^2~\delta({{\mathbf{q}}}_{{\parallel}}-{{\mathbf{q}}}^{\prime}_{{\parallel}})} &=&C^{\mathrm{NE}}_{zz}({{\mathbf{q}}}_{{\parallel}},{z},z^\prime),\\
\frac{{\left\langle \delta\omega^*_z({\mathbf{q}}_{\parallel },z)~\delta\omega_z({\mathbf{q}}^{\prime}_{\parallel},z^\prime)\right\rangle }_{\mathrm{NE}}}{{(2\pi)}^2~ \delta({{\mathbf{q}}}_{{\parallel}}-{{\mathbf{q}}}^{\prime}_{{\parallel}})}&=&W^{\mathrm{NE}}_{zz}({{\mathbf{q}}}_{{\parallel}},{z},z^\prime),\\
\frac{{\left\langle\delta{v}^*_z({\mathbf{q}}_{\parallel },z)~\delta\omega_z({\mathbf{q}}^{\prime}_{\parallel},z^\prime)\right\rangle }_{\mathrm{NE}}}{{(2\pi)}^2~ \delta({{\mathbf{q}}}_{{\parallel}}-{{\mathbf{q}}}^{\prime}_{{\parallel}})}&=&B^{\mathrm{NE}}_{zz}({{\mathbf{q}}}_{{\parallel}},{z},z^\prime).
\end{eqnarray}
From the expressions above we can obtain the correlation functions for $\delta v_x$ and $\delta v_y$ by noting that
\begin{eqnarray}
\delta v_x({\mathbf{q}}_{\parallel },z)=\frac{\mathrm{i}}{q^2_{\parallel }}\left[q_x~\partial_z\delta v_z({\mathbf{q}}_{\parallel },z)+q_y~\delta {\omega }_z({\mathbf{q}}_{\parallel },z)\right],\\
\delta v_y({\mathbf{q}}_{\parallel },z)=\frac{\mathrm{i}}{q^2_{\parallel }}\left[q_y~\partial_z\delta v_z({\mathbf{q}}_{\parallel },z)-q_x~\delta {\omega }_z({\mathbf{q}}_{\parallel },z)\right],
\end{eqnarray}
so that
\begin{eqnarray}\label{ES24}
\frac{{\left\langle \delta v^*_x({\mathbf{q}}_{\parallel },z)~\delta v_x({\mathbf{q}}^{\prime}_{\parallel},z^\prime)\right\rangle }_{\mathrm{NE}}}{{(2\pi)}^2~ \delta({{\mathbf{q}}}_{{\parallel}}-{{\mathbf{q}}}^{\prime}_{{\parallel}})}&=&C^{\mathrm{NE}}_{xx}({{\mathbf{q}}}_{{\parallel}},{z},z^\prime),\\
\frac{{\left\langle \delta{v}^*_y({\mathbf{q}}_{\parallel },z)~\delta{v}_y({\mathbf{q}}^{\prime}_{\parallel},z^\prime)\right\rangle }_{\mathrm{NE}}}{{(2\pi)}^2~ \delta({{\mathbf{q}}}_{{\parallel}}-{{\mathbf{q}}}^{\prime}_{{\parallel}})}&=&C^{\mathrm{NE}}_{yy}({{\mathbf{q}}}_{{\parallel}},{z},z^\prime),
\end{eqnarray}
with
\begin{equation}
\begin{split}
C^{\text{NE}}_{xx}(z,z^\prime)=\frac{q_x^2}{q_{\parallel}^4}&~\partial_z\partial_{z^\prime}C^{\text{NE}}_{zz}(z,z^\prime)+\frac{q_y^2}{q_{\parallel}^4}~W^{\text{NE}}_{zz}(z,z^\prime)\\ &+\frac{q_xq_y}{q_{\parallel}^4} X^{\text{NE}}(z,z^\prime),\\
C^{\text{NE}}_{yy}(z,z^\prime)=\frac{q_y^2}{q_{\parallel}^4}&~\partial_z\partial_{z^\prime}C^{\text{NE}}_{zz}(z,z^\prime)+\frac{q_x^2}{q_{\parallel}^4}~W^{\text{NE}}_{zz}(z,z^\prime)\\ &-\frac{q_xq_y}{q_{\parallel}^4} X^{\text{NE}}(z,z^\prime),
\end{split}\raisetag{18pt}\label{E130}
\end{equation}
where we have introduced
\begin{equation*}
X^{\text{NE}}(z,z^\prime)=
\left[\partial_zB^{\text{NE}}_{zz}(z,z^\prime) +\partial_{z^\prime}B^{\text{NE}*}_{zz}(z^\prime,z)  \right].
\end{equation*}
In the expressions above for simplicity we do not specify explicitly the dependence of the various functions on the wave vector $\mathbf{q}_\parallel$.

To solve the fluctuating hydrodynamics equations for $\delta{v}_z(\omega,\mathbf{q}_\parallel,z)$ and $\delta{\omega}_z(\omega,\mathbf{q}_\parallel,z)$ we have adopted a Galerkin approximation. Specifically, to satisfy the boundary conditions, we assume that~\cite{miCouette,miCouette2}
\begin{equation}\label{E125}
\begin{split}
\delta{v}_z(z)&= (z^2-1)^2\left[A_0+A_1~z+A_2~z^2+\cdots\right],\\
\delta{\omega}_z(z)&= (z^2-1)\left[B_0+B_1~z+B_2~z^2+\cdots\right],\\
\end{split}
\end{equation}
where the coefficients $A_N(\omega,\mathbf{q}_\parallel)$ and $B_N(\omega,\mathbf{q}_\parallel)$ are determined by projection of the equations onto the basis used for the expansion~\eqref{E125} itself, \emph{i.e.}, $(z^2-1)^2~z^N$ and $(z^2-1)~z^N$, and solving the resulting algebraic equations. These coefficients depend on the two-dimensional wave vector $\mathbf{q}_\parallel$ and the frequency $\omega$ of the fluctuations, which was not explicitly indicated in Eqs.~\eqref{E125} above for the sake of simplicity.

In practice, the Galerkin approach is only useful when the expansion~\eqref{E125} is truncated at some low order, and we have truncated at $N=1$. As before and for illustrative purposes, we consider in detail here only $V_{zz}^0$. For the nonequilibrium contribution $C_{zz}^\text{NE}(\mathbf{q}_\parallel,z)$ in Eq.~\eqref{ES17} the first-order Galerkin approximation described above, after substitution of $z=z^\prime$ and integration over $\omega$, yields:
\begin{equation}\label{E41}
C_{zz}^\text{NE}(z)=\tilde{S}{Re} \left(1-z^2\right)^4 \left[C_1 - C_2~z^2\right]~{Re}^2.
\end{equation}
Generally, the coefficients $C_{1}(\text{Re},\mathbf{q}_\parallel)$ and $C_{2}(\text{Re},\mathbf{q}_\parallel)$ in the equation above depend on both the Reynolds number and on the two-dimensional wave vector $\mathbf{q}_\parallel$ of the fluctuations. However, since only the limit $\text{Re}\ll1$ of these expressions shall be used later, we only explicitly show their expressions for $\text{Re}=0$, namely
\begin{equation}\label{E42}
\hspace*{-12pt}C_{1}(0,\mathbf{q}_\parallel)=\dfrac{3465}{128}\frac{\dfrac{q_x^2~q_\parallel^2(11+q_\parallel^2)}{(63+12q_\parallel^2+2q_\parallel^4)}} {11(1089+411q_\parallel^2+42q_\parallel^4+2q_\parallel^6)},
\end{equation}
and
\begin{equation}
C_{2}(0,\mathbf{q}_\parallel)=\dfrac{3465}{128}\frac{\dfrac{q_x^2~q_\parallel^4}{(495+44q_\parallel^2+2q_\parallel^4)}}{1089+411q_\parallel^2+42q_\parallel^4+2q_\parallel^6}.
\end{equation}
Next, we can define a $z$-dependent component $V^0_{zz}(z)$ which, in terms of the dimensionless units adopted in this section, will be given by the expression:
\begin{equation}\label{ES29}
V^0_{zz}(z)=\displaystyle\lim_{\mathrm{Re}\ll1}~ \frac{1}{\tilde{S}{Re}^3} \frac{1}{(2\pi)^2}\int_{\mathbb{R}^2} C_{zz}^\text{(NE)}(\mathbf{q}_\parallel,z)~d\mathbf{q}_\parallel.
\end{equation}
Upon substitution of Eq.~\eqref{E41} into Eq.~\eqref{ES29} and considering the $\text{Re}\ll1$ limit, we obtain:
\begin{equation}
\begin{split}
V^0_{zz}(z)=\frac{1}{(2\pi)^2}\left(1-z^2\right)^4& \left[\int_{\mathbb{R}^2}C_1(0,\mathbf{q}_\parallel)~d\mathbf{q}_\parallel\right.\\
&\left. - z^2 \int_{\mathbb{R}^2} C_2(0,\mathbf{q}_\parallel)~~d\mathbf{q}_\parallel\right].
\end{split}\raisetag{50pt}\label{ES30}
\end{equation}
The two integrals are convergent and can be performed analytically, but the result is long and not particularly informative. We prefer to display the result numerically, namely
\begin{equation}\label{E4}
V_{zz}^{0}(z)=\left(1-z^2\right)^4\left(1.584-6.812~z^2 \right)\times10^{-3}.
\end{equation}
As explained in Section~\ref{S2}, the quantity relevant for the estimation of fluctuation-induced pressures is obtained upon averaging over the wall-normal coordinate, or
\begin{equation}\label{E51}
V_{zz}^{0}=\frac{1}{2}\int_{-1}^1 V_{zz}^{0}(z)~dz=~3.92\times10^{-4},
\end{equation}
which is the value quoted in the main text. The other coefficients, $V_{xx}^0$ and $V_{yy}^0$, are evaluated in a similar fashion from Eq.~\eqref{E130}. The resulting values are $V^0_{xx}=+0.001438$ and $V^0_{yy}=+0.000480$. Further discussion of these results and comparison with the large Re results was presented in Section~\ref{S3}.


%

\end{document}